\documentclass[iop,apjl,natbib]{emulateapj}
\usepackage{graphicx}
\usepackage{color}

\newcommand{\V}[1]{\mathbf{#1}} 
\newcommand{\T}[1]{\texttt{#1}} 
\newcommand\Alfven{Alfv\'en }

\newcommand{\eqref}[1]{Eq.~(\ref{#1})}
\newcommand{\xhat}{\mbox{$\hat{\mathbf{x}}$}} 
\newcommand{\yhat}{\mbox{$\hat{\mathbf{y}}$}} 
\newcommand{\zhat}{\mbox{$\hat{\mathbf{z}}$}}

\begin{document}

\title{The Dynamical Generation of Current Sheets in Astrophysical 
Plasma Turbulence}

\author{Gregory G.~Howes}

\affil{Department of Physics and Astronomy, University of Iowa, Iowa City, IA 52242, USA}

\begin{abstract}
Turbulence profoundly affects particle transport and plasma heating in
many astrophysical plasma environments, from galaxy clusters to the
solar corona and solar wind to Earth's magnetosphere. Both fluid and
kinetic simulations of plasma turbulence ubiquitously generate
coherent structures, in the form of current sheets, at small scales,
and the locations of these current sheets appear to be associated with
enhanced rates of dissipation of the turbulent energy. Therefore,
illuminating the origin and nature of these current sheets is critical
to identifying the dominant physical mechanisms of dissipation, a
primary aim at the forefront of plasma turbulence research. Here we
present evidence from nonlinear gyrokinetic simulations that strong
nonlinear interactions between counterpropagating \Alfven waves, or
strong \Alfven wave collisions, are a natural mechanism for the
generation of current sheets in plasma turbulence. Furthermore, we
conceptually explain this current sheet development in terms of the
nonlinear dynamics of \Alfven wave collisions, showing that these
current sheets arise through constructive interference among the
initial \Alfven waves and nonlinearly generated modes.  The properties
of current sheets generated by a strong \Alfven wave collisions are
compared to published observations of current sheets in the Earth's
magnetosheath and the solar wind, and the nature of these current
sheets leads to the expectation that Landau damping of the constituent
\Alfven waves plays a dominant role in the damping of turbulently
generated current sheets.
\end{abstract}



\maketitle 

\section{Introduction}
The ubiquitous presence of turbulence impacts the evolution of many
space and astrophysical plasma environments, mediating the transport
of energy from violent events or instabilities at large scales down to
the small scales at which the energy is ultimately converted to heat
of the protons, electrons, and minor ions.  It is widely believed that
plasma turbulence plays an important role in heating the solar corona
to millions of degrees Kelvin, accelerating the solar wind to hundreds
of kilometers per second, regulating star formation, transporting heat
in galaxy clusters, and affecting the injection of particles and
energy into the Earth's magnetosphere.  At the forefront of plasma
turbulence research is the effort to identify the physical mechanisms
by which the turbulent fluctuations are damped and their energy
converted to plasma heat or some other energization of particles.

In contrast to the intermittent filaments of vorticity that arise in
hydrodynamic turbulence \citep{She:1990}, intermittent current sheets
are found to develop in plasma turbulence
\citep{Matthaeus:1980,Meneguzzi:1981}. Recent work investigating the
statistics of these coherent structures, self-consistently generated
by the plasma turbulence, has demonstrated that the dissipation of
turbulent energy is largely concentrated in these current sheets
\citep{Uritsky:2010,Osman:2011,Zhdankin:2013}. Since current sheets
are associated with enhanced dissipation, illuminating their origin
and nature is critical to identifying the dominant physical mechanisms
of dissipation in plasma turbulence.

How current sheets develop in plasma turbulence is a longstanding
question in the study of space and astrophysical plasmas
\citep{Parker:1972,Pouquet:1978,Priest:1985,vanBallegooijen:1985,Antiochos:1987,Zweibel:1987,Biskamp:1989,Longcope:1994,Cowley:1997,Spangler:1999,Biskamp:2000,Merrifield:2005,Greco:2008}.
Early work focused on the study of  solar coronal loops, asking if
the continuous motion of line-tied footpoints in ideal MHD would lead
an initially smooth magnetic field to develop a tangential
discontinuity \citep{Parker:1972}, necessarily supported by a sheet of
finite current, according to Maxwell's equations. It was argued that
continuous footpoint motion cannot generate a discontinuous magnetic
field \citep{vanBallegooijen:1985,Antiochos:1987,Zweibel:1987}, but
later shown that current layers of finite but arbitrarily small
thickness were realizable through continuous footpoint motion
\citep{Longcope:1994,Cowley:1997}.  Of course, in the more complete
kinetic plasma description, current layers generally have structure at
both characteristic ion and electron length scales; kinetic
simulations of plasma turbulence indeed observe the development of
current sheets of finite thickness
\citep{Wan:2012,Karimabadi:2013,TenBarge:2013a}.

Recent spacecraft measurements of current sheets in the near-Earth
solar wind have lead to fundamental questions about their origin and
their influence on plasma heating. Do the measured current sheets
represent advected flux tube boundaries
\citep{Borovsky:2008,Borovsky:2010}, or are they generated dynamically
by the turbulence itself \citep{Boldyrev:2011,Zhdankin:2012}?  In the
past few years, vigorous activity has focused on the spatial
localization of plasma heating by the dissipation of turbulence in
current sheets through statistical analyses of solar wind observations
\citep{Osman:2011,Borovsky:2011,Osman:2012a,Perri:2012a,Wang:2013,Wu:2013,Osman:2014b}
and numerical simulations
\citep{Wan:2012,Karimabadi:2013,TenBarge:2013a,Wu:2013,Zhdankin:2013}.

To understand the origin of coherent structures, we must investigate
how the turbulent nonlinear interactions govern their development
\citep{Howes:2015b}. In plasma turbulence, the \Alfven wave represents the
fundamental response of the plasma to an applied perturbation.  Early
research on incompressible MHD turbulence in the 1960s
\citep{Iroshnikov:1963,Kraichnan:1965} emphasized the wave-like nature
of turbulent plasma motions, suggesting that nonlinear interactions
between counterpropagating \Alfven waves---or \emph{\Alfven wave
  collisions}---mediate the turbulent cascade of energy from large to
small scales.  The \Alfven wave remains central to modern theories of
MHD turbulence that provide explanations for the anisotropic nature of
the turbulent cascade \citep{Goldreich:1995} and the dynamic alignment
of velocity and magnetic field fluctuations \citep{Boldyrev:2006}.

In this Letter, we demonstrate that the generation of current sheets
in plasma turbulence is a natural consequence of strong \Alfven wave
collisions. Furthermore, we present a first-principles explanation for
this current sheet development in terms of the nonlinear dynamics,
showing that the current sheet can be accurately reconstructed from a
linear superposition of the interacting \Alfven waves and a
surprisingly small number of nonlinearly generated modes. The
properties of the resulting current sheet are compared to previously
published observations of current sheets in turbulence in the Earth's
magnetosheath \citep{Retino:2007,Sundkvist:2007} and the solar wind
\citep{Perri:2012a}. Finally, we discuss implications for the damping
of current sheets generated by strong \Alfven wave collisions.

\section{Alfv\'en wave collisions}
Although the incompressible MHD equations lack a number of physical
effects that occur in realistic space and astrophysical plasmas, they
do contain the minimal ingredients that lead to the anisotropic
cascade and development of current sheets in plasma turbulence.
Expressed here in the symmetric Elsasser form, these equations are
\begin{equation}
\frac{\partial \V{z}^{\pm}}{\partial t} 
\mp \V{v}_A \cdot \nabla \V{z}^{\pm} 
=-  \V{z}^{\mp}\cdot \nabla \V{z}^{\pm} -\nabla P/\rho_0,
\label{eq:elsasserpm}
\end{equation}
and $\nabla\cdot \V{z}^{\pm}=0$.  Here $\V{v}_A =\V{B}_0/\sqrt{4
  \pi\rho_0}$ is the \Alfven velocity due to the equilibrium field
$\V{B}_0=B_0 \zhat$ where $\V{B}=\V{B}_0+ \delta \V{B} $, $P$ is total
pressure (thermal plus magnetic), $\rho_0$ is mass density, and
$\V{z}^{\pm} = \V{u} \pm \delta \V{B}/\sqrt{4 \pi \rho_0}$ are the
Elsasser fields which represent waves that propagate up or down the
mean magnetic field. The nonlinear term, $\V{z}^{\mp}\cdot \nabla
\V{z}^{\pm} $, governs the nonlinear interactions between
counterpropagating \Alfven waves, or \Alfven wave collisions.  The
\emph{nonlinearity parameter}, the ratio of the nonlinear to the
linear term magnitudes in eq.~(\ref{eq:elsasserpm}), $\chi \equiv
|\V{z}^{\mp}\cdot \nabla \V{z}^{\pm}|/|\V{v}_A \cdot \nabla
\V{z}^{\pm}|$, characterizes the strength of the nonlinearity.  The
limit of strong incompressible MHD turbulence occurs when $\chi \sim
1$, in which the nonlinear energy transfer timescale is comparable to
the linear wave period, a condition known as critical balance
\citep{Goldreich:1995}.

Following significant previous studies on weak incompressible MHD
turbulence \citep{Sridhar:1994,Ng:1996,Galtier:2000}, the nonlinear
energy transfer in \Alfven wave collisions has recently been solved
analytically in the weakly nonlinear limit $\chi \ll 1$
\citep{Howes:2013a}, confirmed numerically with gyrokinetic numerical
simulations \citep{Nielson:2013a}, and verified experimentally in the
laboratory \citep{Howes:2012b}, establishing \Alfven wave collisions as
the fundamental building block of astrophysical plasma turbulence.
Here we briefly review the details of the nonlinear energy transfer
in the weak turbulence limit, $\chi \ll\ 1$.

\begin{figure}
\begin{center}
\resizebox{3.2in}{!}{\includegraphics*[0.32in,2.25in][8.2in,9.75in]{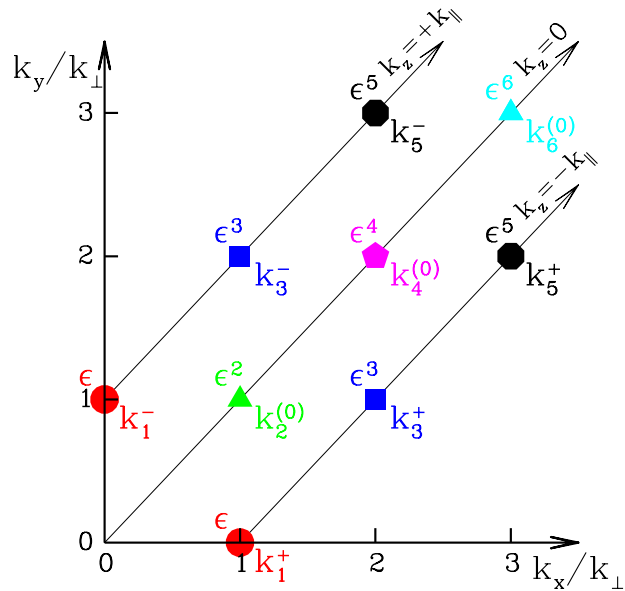}}
\end{center}
\caption{Perpendicular Fourier modes nonlinearly generated by an
  \Alfven wave collision between initial counterpropagating \Alfven
  waves $\V{k}_1^+$ and $\V{k}_1^-$.
\label{fig:modes_o5}}
\end{figure}

Consider the nonlinear interaction between two perpendicularly
polarized, counterpropagating \Alfven waves with wavevectors
$\V{k}_1^+ = k_\perp \xhat - k_\parallel \zhat$ and $\V{k}_1^- =
k_\perp \yhat + k_\parallel \zhat$, as shown in
Fig.~\ref{fig:modes_o5} (red circles), where $k_\perp$ and
$k_\parallel$ are positive constants. The lowest order nonlinear
interaction in the asymptotic solution creates an inherently
nonlinear, purely magnetic mode with wavevector $\V{k}_2^{(0)}=
k_\perp \xhat + k_\perp \yhat$ (green triangle).  This secondary mode
has $k_z=0$ and frequency $\omega=2 k_\parallel v_A$, but does not
grow secularly in time.  At next order, the primary modes then
interact with this secondary $k_z=0$ mode to transfer energy secularly
to two nonlinearly generated \Alfven waves (blue squares), where
$\V{k}_1^+ $ transfers energy to an \Alfven wave with $\V{k}_3^+ = 2
k_\perp \xhat + k_\perp \yhat - k_\parallel \zhat$, and $\V{k}_1^-$
transfers energy to $\V{k}_3^- = k_\perp \xhat+ 2k_\perp \yhat +
k_\parallel \zhat$. This process is the fundamental mechanism by which
turbulence transfers energy anisotropically from large to small scales
\citep{Howes:2013a}.

\section{Strong Alfv\'en Wave Collision Simulations}
Although the analytical solution for the dynamics of \Alfven wave
collisions is calculated in the MHD approximation, we employ here a
gyrokinetic code for two reasons: (1) to demonstrate that the physical
mechanism is not altered under the weakly collisional plasma
conditions relevant to many space and astrophysical plasma
environments; and, (2) to enable the direct comparison of current
sheet profiles from our simulations to spacecraft measurements in the
weakly collisional solar wind. We employ the Astrophysical
Gyrokinetics code \T{AstroGK} \citep{Numata:2010} to perform a
gyrokinetic simulation of the nonlinear interaction between two
counterpropagating \Alfven waves in the strongly nonlinear limit,
referred to as a \emph{strong \Alfven wave collision}.

\begin{figure*}
\hbox{
 \hfill
 \resizebox{2.3in}{!}{\includegraphics{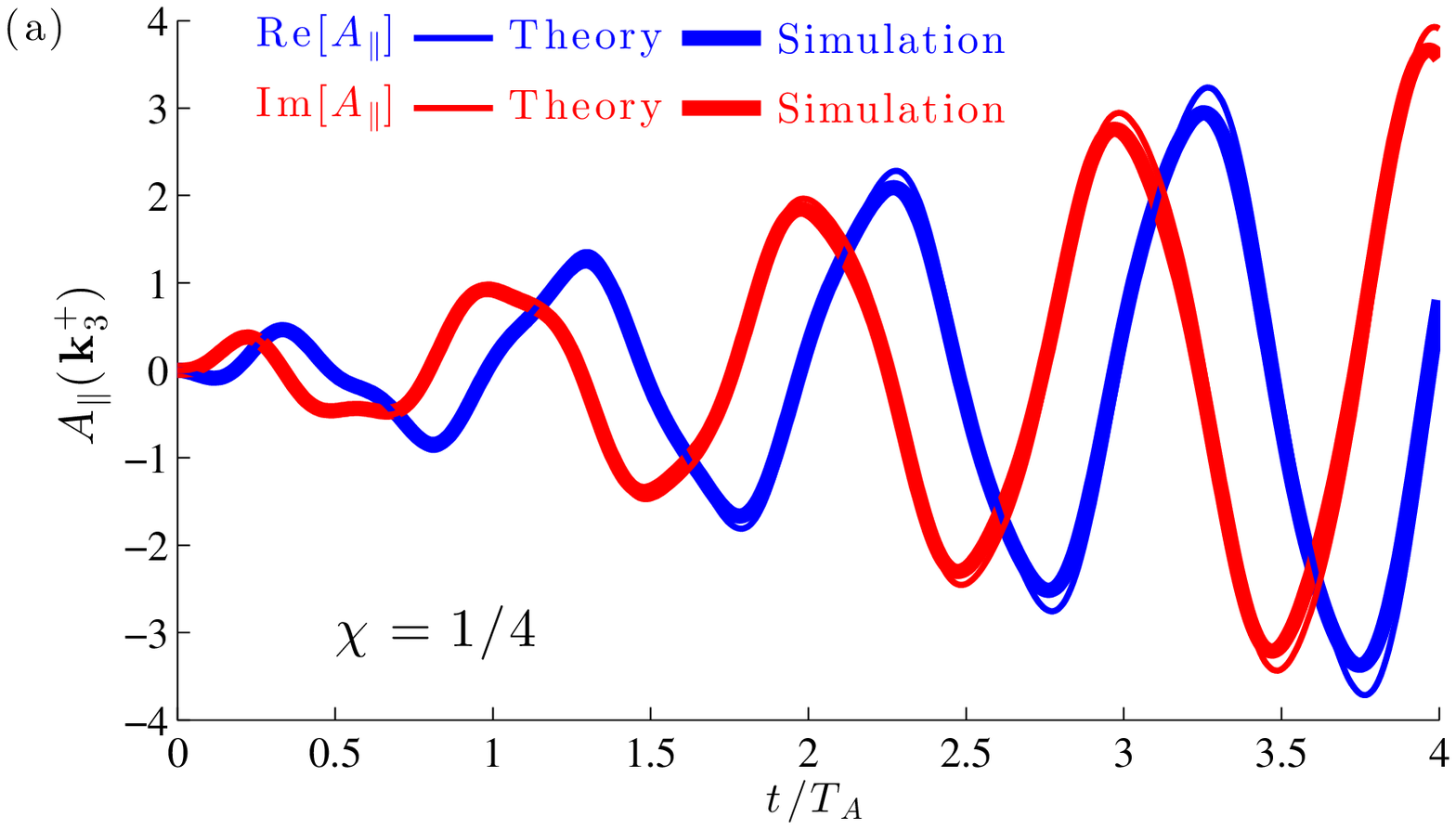}}
 \hfill
 \resizebox{2.3in}{!}{\includegraphics{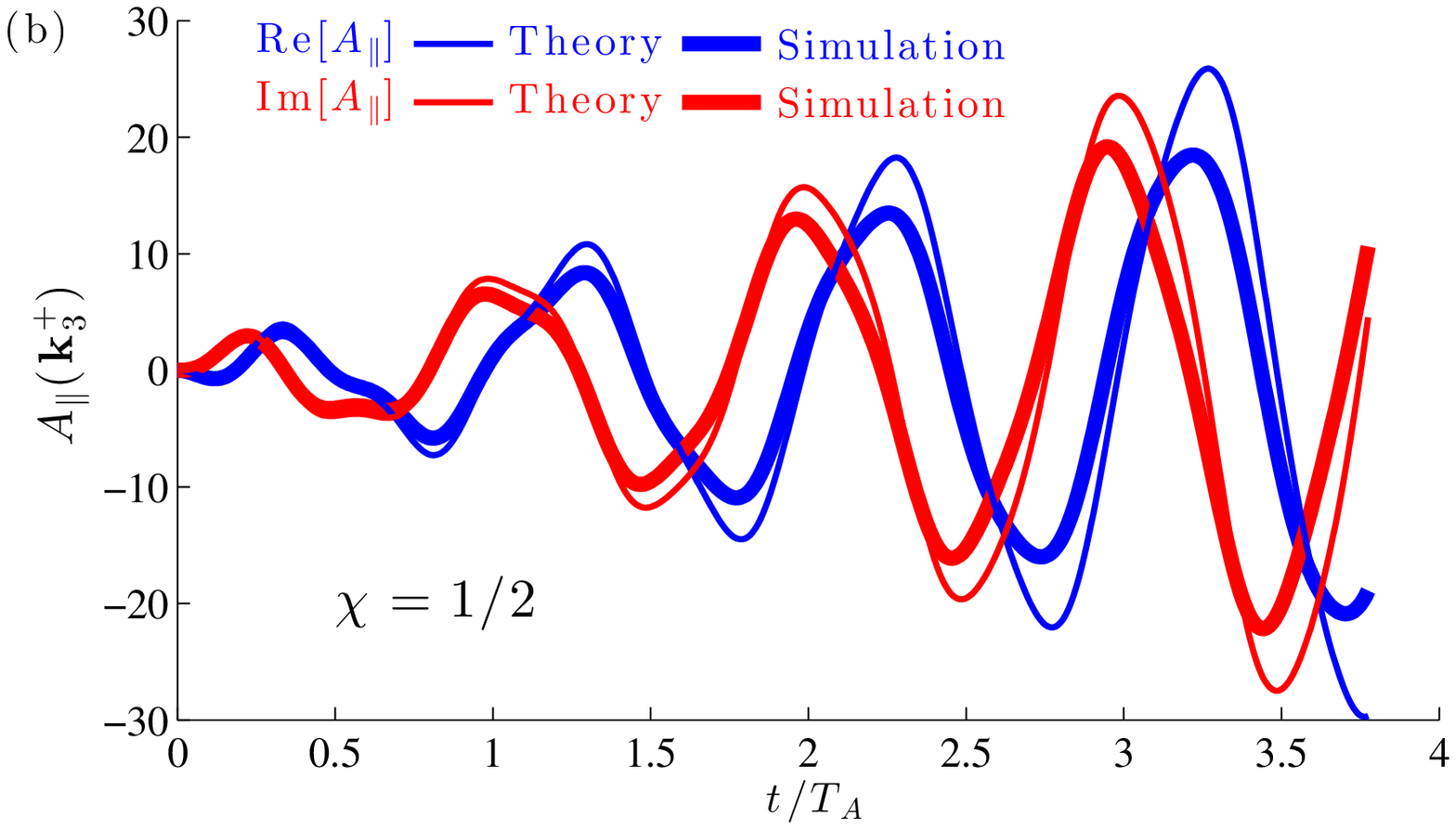}}
 \hfill
 \resizebox{2.3in}{!}{\includegraphics{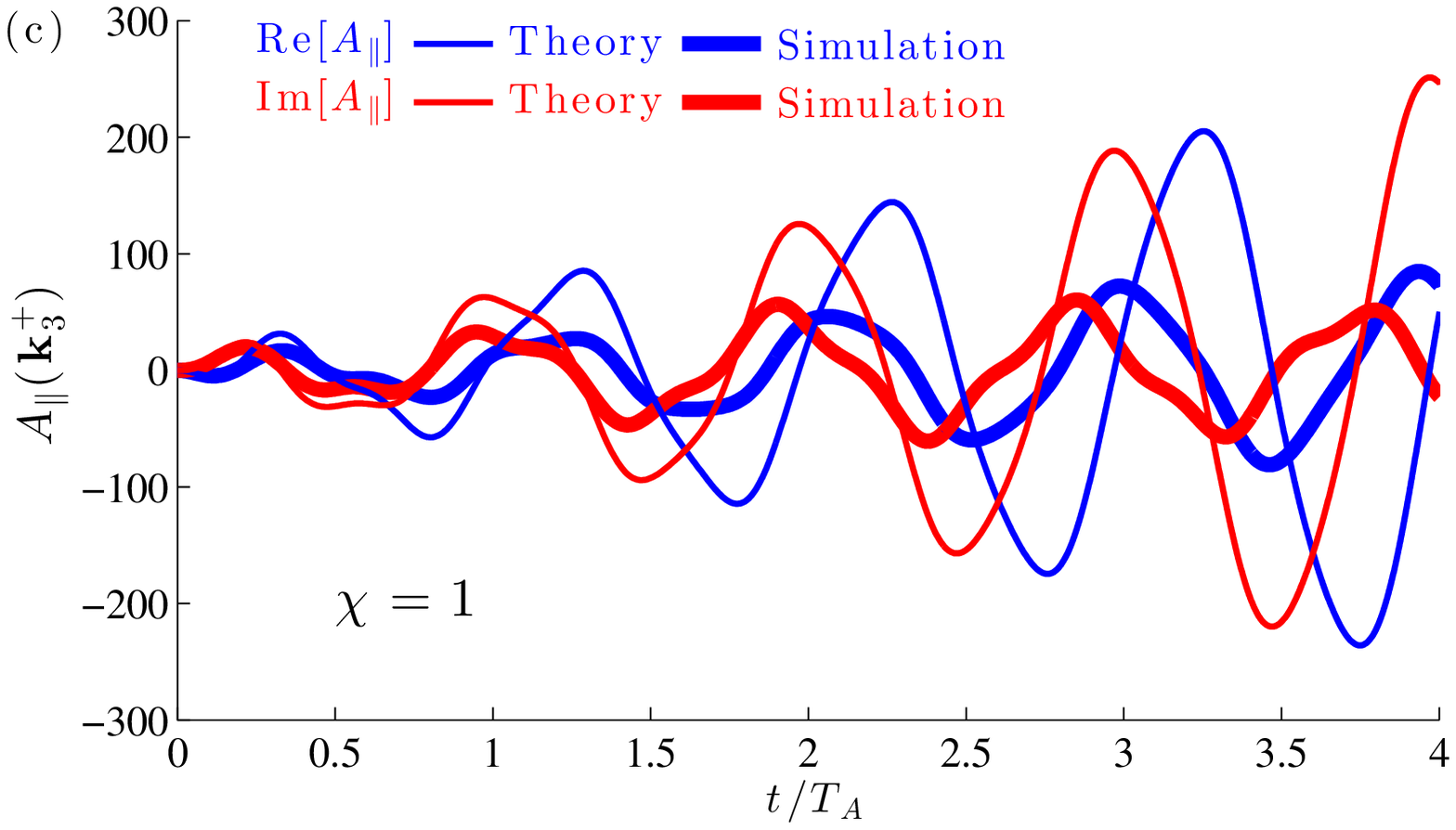}}
 \hfill }
\caption{Comparison of the real (blue) and imaginary
  (red) parts of the complex variable $A_\parallel(\V{k}_3^+)$ between
  the predictions of the asymptotic analytical solution (thin) and
  nonlinear gyrokinetic numerical simulations (thick) with increasing
  nonlinearity parameters, from moderately strong nonlinearity with
  (a) $\chi=1/4$, through (b) $\chi=1/2$, to strong nonlinearity with
  (c) $\chi=1$.
\label{fig:chi_scan}}
\end{figure*}

\T{AstroGK} evolves the perturbed gyroaveraged distribution function
$h_s(x,y,z,\lambda,\varepsilon)$ for each species $s$, the scalar
potential $\varphi$, the parallel vector potential $A_\parallel$, and the
parallel magnetic field perturbation $\delta B_\parallel$ according to
the gyrokinetic equation and the gyroaveraged Maxwell's equations\citep{Frieman:1982,Howes:2006}. Velocity space coordinates are
$\lambda=v_\perp^2/v^2$ and $\varepsilon=v^2/2$. The domain is a
periodic box of size $L_{\perp }^2 \times L_{\parallel }$, elongated
along the straight, uniform mean magnetic field $\V{B}_0=B_0 \zhat$,
where all quantities may be rescaled to any parallel dimension
satisfying $L_{\parallel } /L_{\perp } \gg 1$. Uniform Maxwellian
equilibria for ions (protons) and electrons are chosen, with the
correct mass ratio $m_i/m_e=1836$. Spatial dimensions $(x,y)$
perpendicular to the mean field are treated pseudospectrally; an
upwind finite-difference scheme is used in the parallel direction,
$z$. Collisions employ a fully conservative,
linearized collision operator with energy diffusion and
pitch-angle scattering \citep{Abel:2008,Barnes:2009}.

To set up the simulation of an \Alfven wave collision, following
\citet{Nielson:2013a}, we initialize two perpendicularly polarized,
counterpropagating plane \Alfven waves, $\V{z}^{+} = z_+ \cos (k_\perp
x -k_\parallel z -\omega_0 t) \yhat$ and $\V{z}^{-} =z_- \cos (k_\perp
y +k_\parallel z -\omega_0 t)\xhat$, where $\omega_0=k_\parallel v_A$,
$k_\perp=2 \pi/L_\perp$, and $k_\parallel=2 \pi/L_\parallel$. We
specify a balanced collision with equal counterpropagating wave
amplitudes, $z_+=z_-$, so the nonlinearity parameter is $\chi=k_\perp
z_\pm/(k_\parallel v_A)$. To study the nonlinear evolution in the
limit $k_\perp \rho_i\ll 1$, we choose a perpendicular simulation
domain size $L_{\perp}=40\pi \rho_i$ with simulation resolution
$(n_x,n_y,n_z,n_\lambda,n_\varepsilon,n_s)= (64,64,64,32,16,2)$.  The
fully resolved perpendicular range in this dealiased pseudospectral
method covers $0.05 \le k_\perp \rho_i \le 1.05$. Here the ion thermal
Larmor radius is $\rho_i= v_{ti}/\Omega_i$, the ion thermal velocity
is $v_{ti}^2 = 2T_i/m_i$, the ion cyclotron frequency is $\Omega_i=
q_i B_0/(m_i c)$, and the temperature is given in energy units. The
plasma parameters, relevant to near-Earth solar wind conditions, are
$\beta_i=1$ and $T_i/T_e=1$.

To demonstrate that the dominant physical
pathway of nonlinear energy transfer, elucidated analytically in the
weak turbulence limit $\chi \ll 1$ \citep{Howes:2013a}, persists as the
turbulence reaches the strong turbulence limit $\chi \rightarrow 1$, we
compare the analytical prediction for the nonlinear evolution of the
parallel vector potential for mode $\V{k}_3^+$,
$A_\parallel(\V{k}_3^+)$, to simulation results with increasingly
strong nonlinearity. In Fig.~\ref{fig:chi_scan}, we plot the real (blue) and imaginary
(red) parts of the complex parallel vector potential
$A_\parallel(\V{k}_3^+)$ for \Alfven wave collision simulations with
(a) $\chi=1/4$, (b) $\chi=1/2$, and (c) $\chi=1$. Although the
analytical solution \citep{Howes:2013a} is strictly valid only for weak
nonlinearity $\chi \ll 1$, even for the moderately strong nonlinearity of
$\chi=1/2$, the solution remains relatively accurate; at $\chi=1$, the
solution ceases to be quantitatively correct (since the primary modes
lose significant energy through the vigorous nonlinear energy
transfer), but the general picture of the secular transfer of energy
mediated by nonlinearly generated $k_z=0$ modes to nonlinearly
produced daughter \Alfven waves remains qualitatively correct.

\begin{figure*}
  \vbox{\hsize 6.6in \hspace*{0.25in} (a) Full Simulation (945 modes)  \hspace{0.55in}  (b) 2 modes\hspace{1.7in}  (c) 3 modes}
\hbox{
  \hfill
\resizebox{2.3in}{!}{\includegraphics{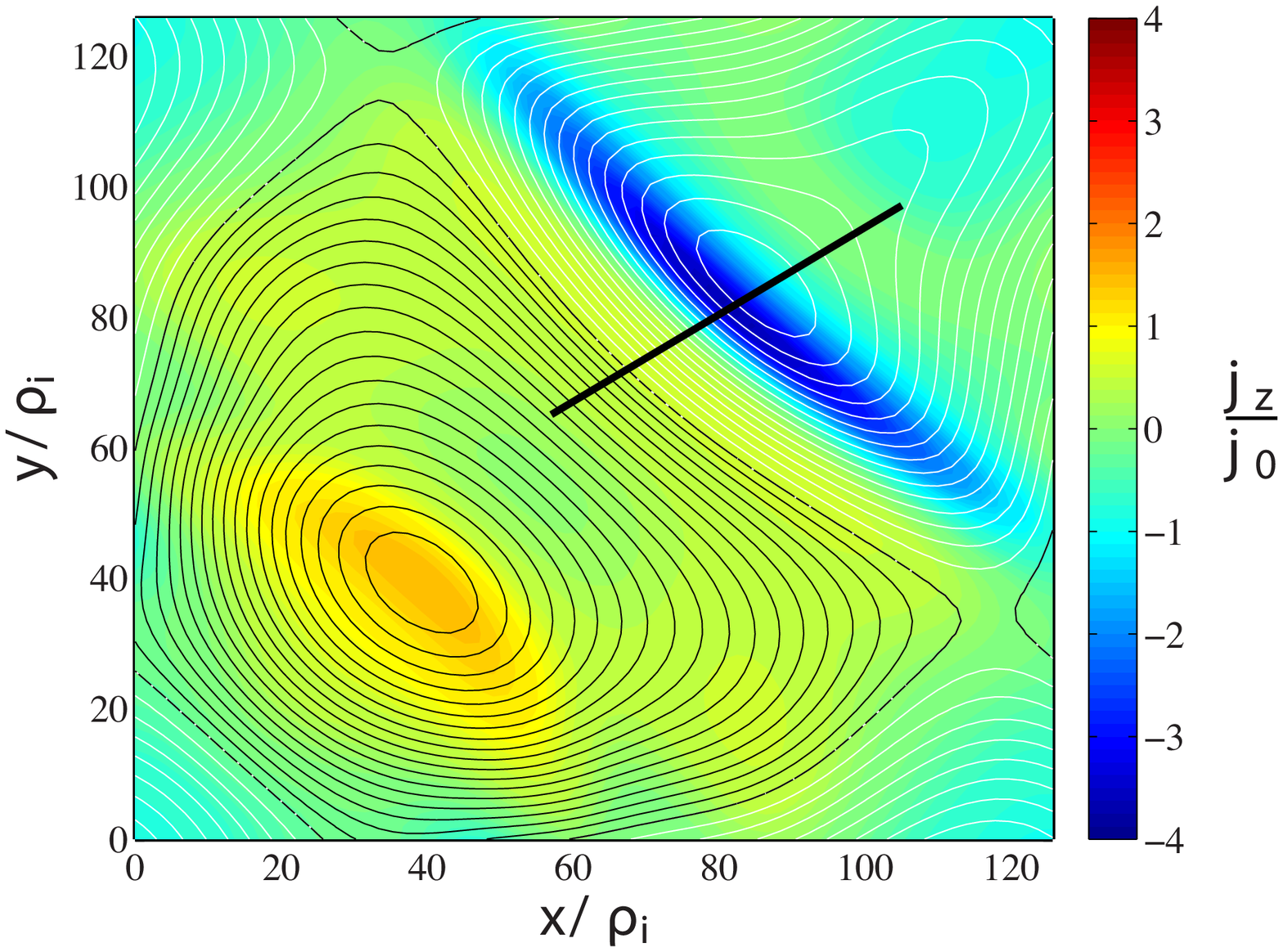}}
  \hfill
\resizebox{2.3in}{!}{\includegraphics*[0.65in,2.6in][8.0in,8.05in]{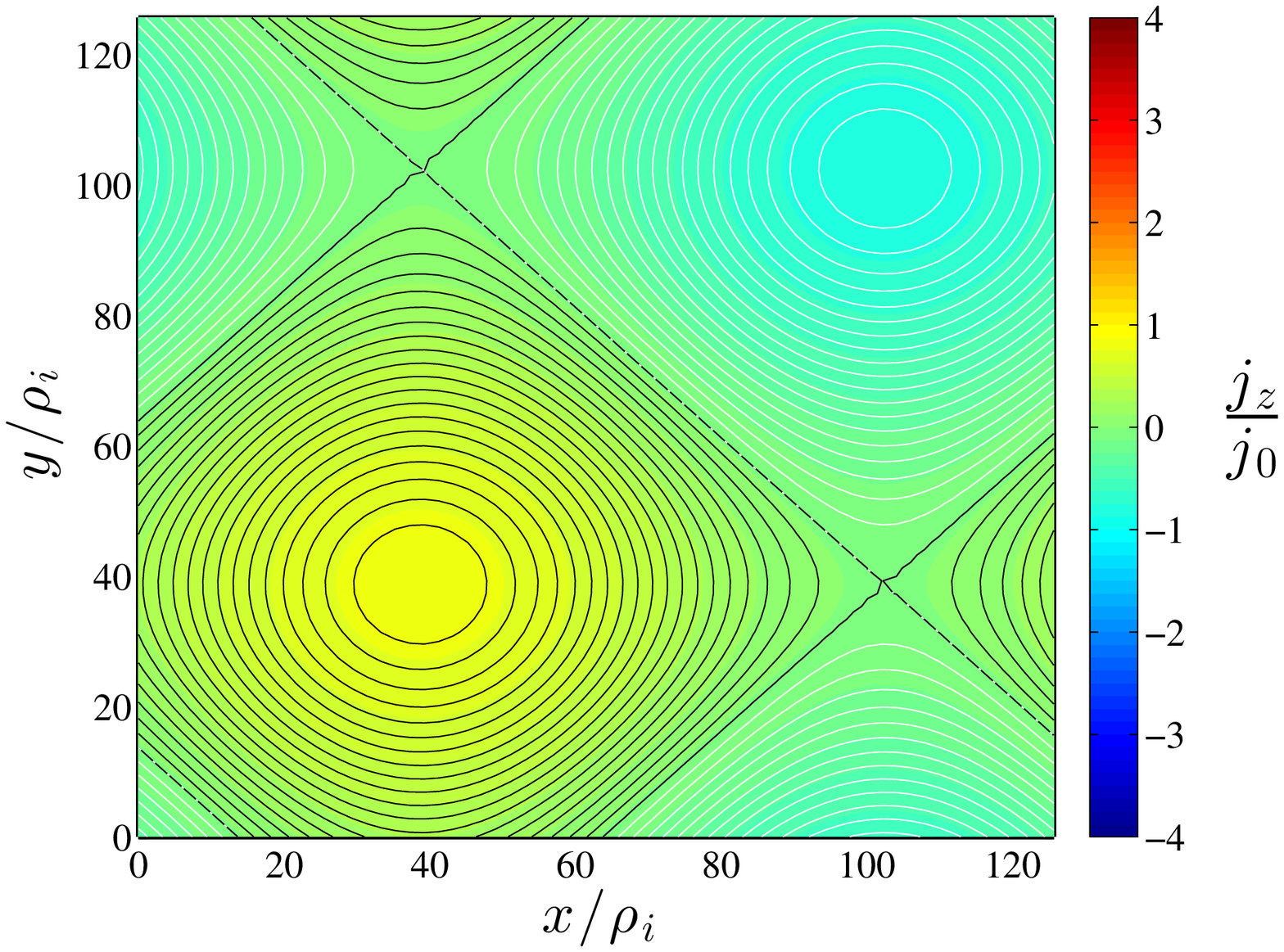}} \hfill
 \hfill
\resizebox{2.3in}{!}{\includegraphics*[0.65in,2.6in][8.0in,8.05in]{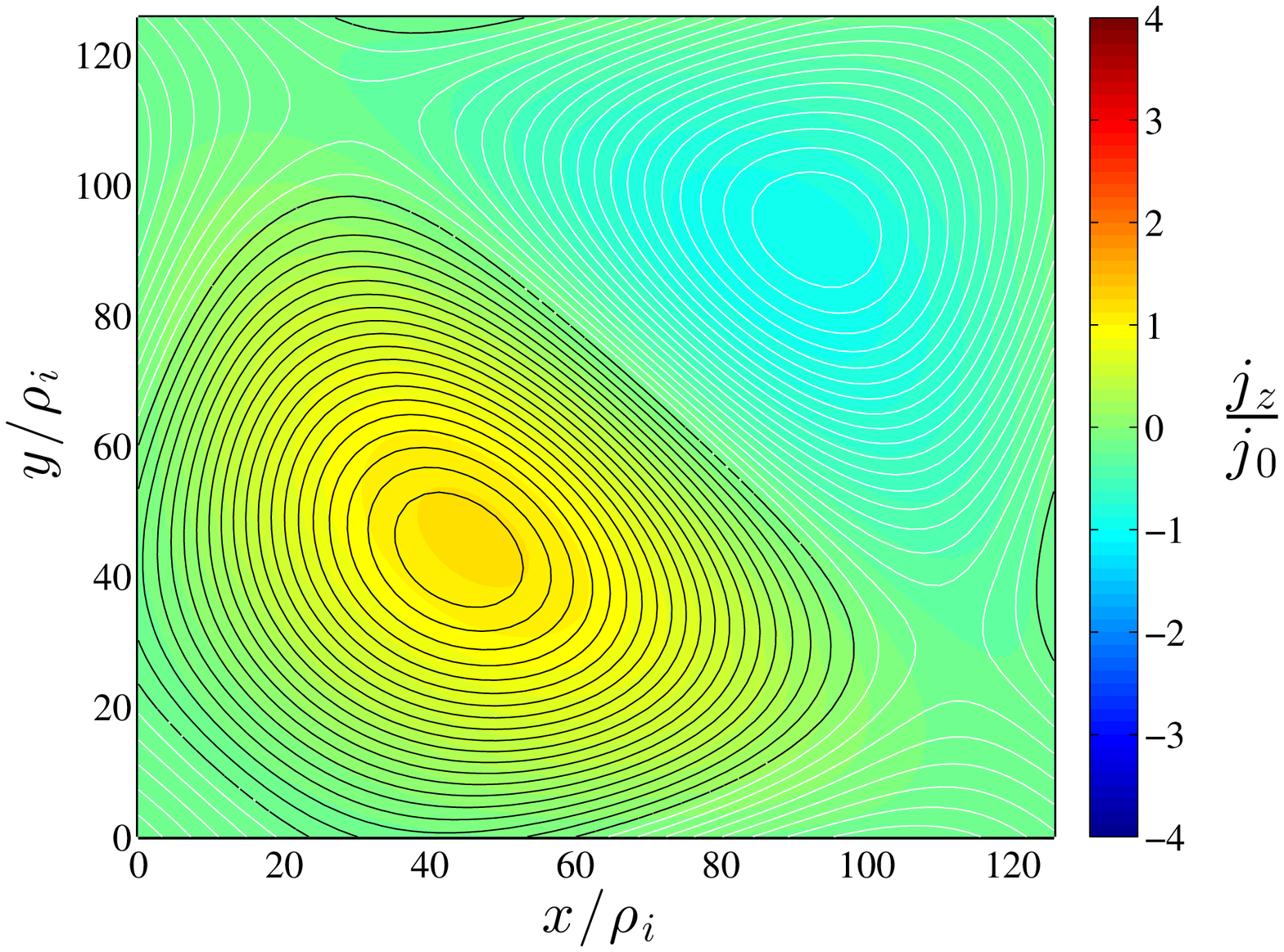}} \hfill}
  \vbox{\hsize 6.6in \hspace*{0.25in} (d) 6 modes  \hspace{1.65in} (e) 9 modes \hspace{1.65in} (f) 12 modes }
\hbox{
 \hfill\resizebox{2.3in}{!}{\includegraphics*[0.65in,2.6in][8.0in,8.05in]{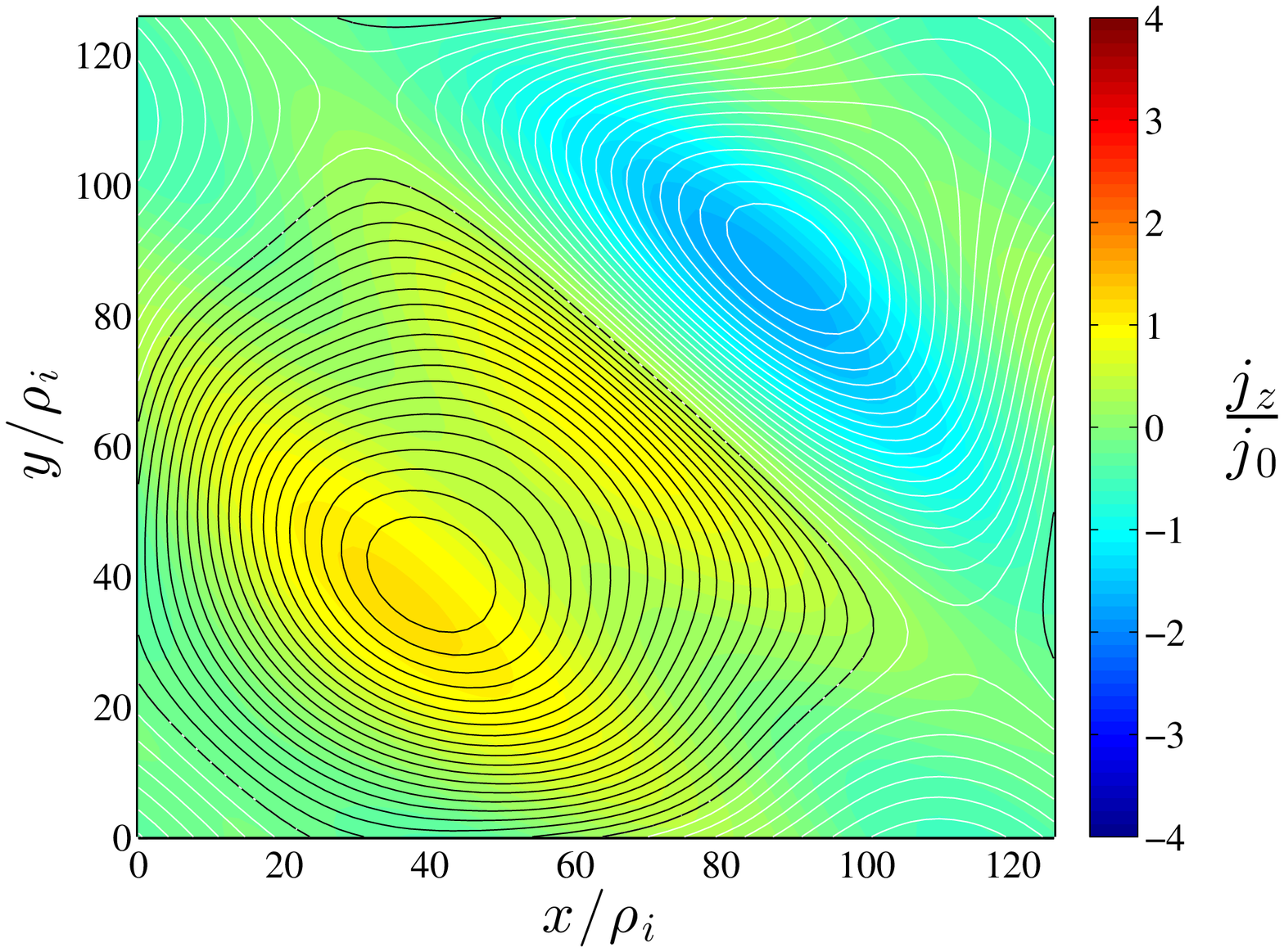}} \hfill
\resizebox{2.3in}{!}{\includegraphics*[0.65in,2.6in][8.0in,8.05in]{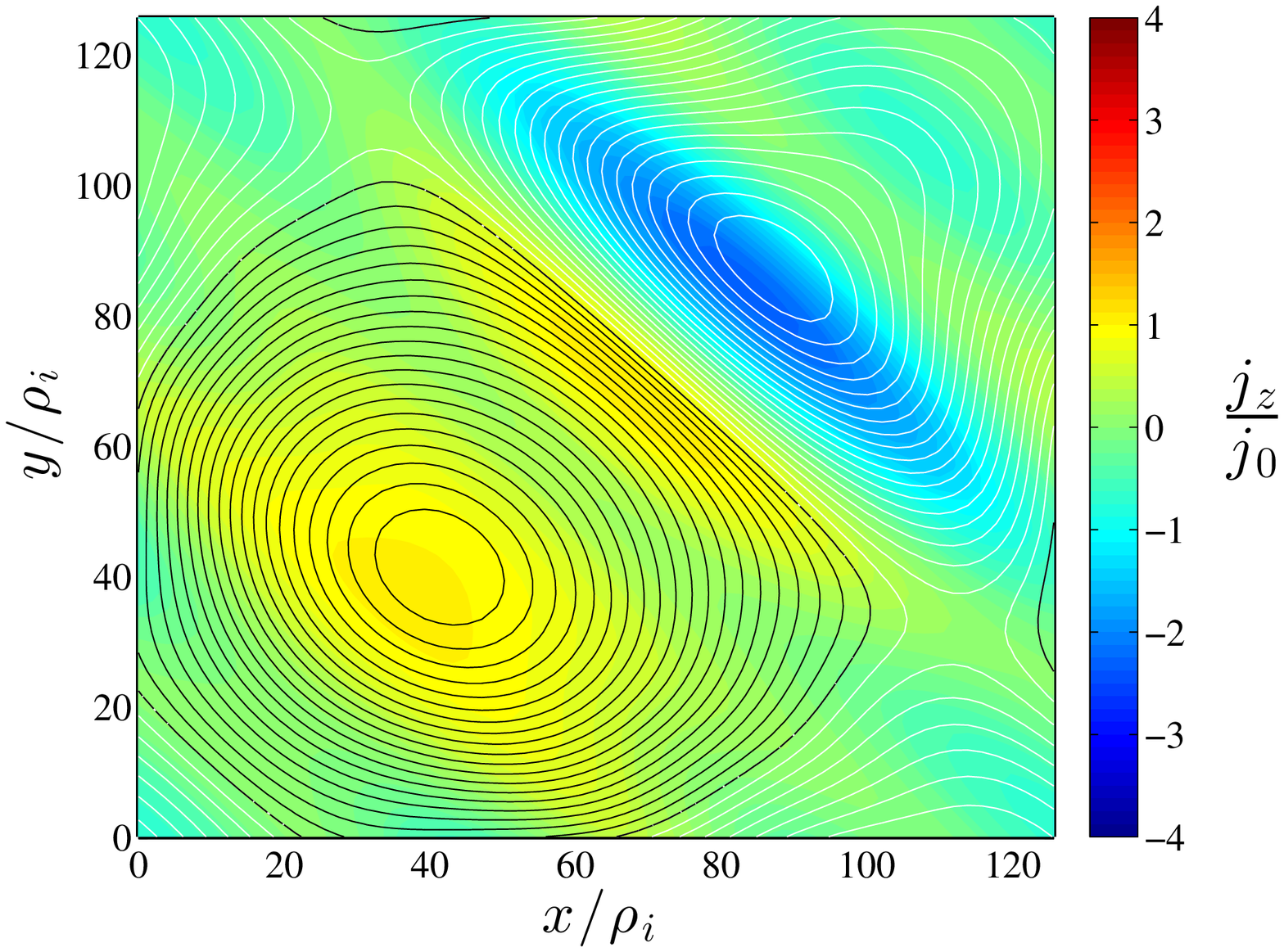}}  \hfill
\resizebox{2.3in}{!}{\includegraphics*[0.65in,2.6in][8.0in,8.05in]{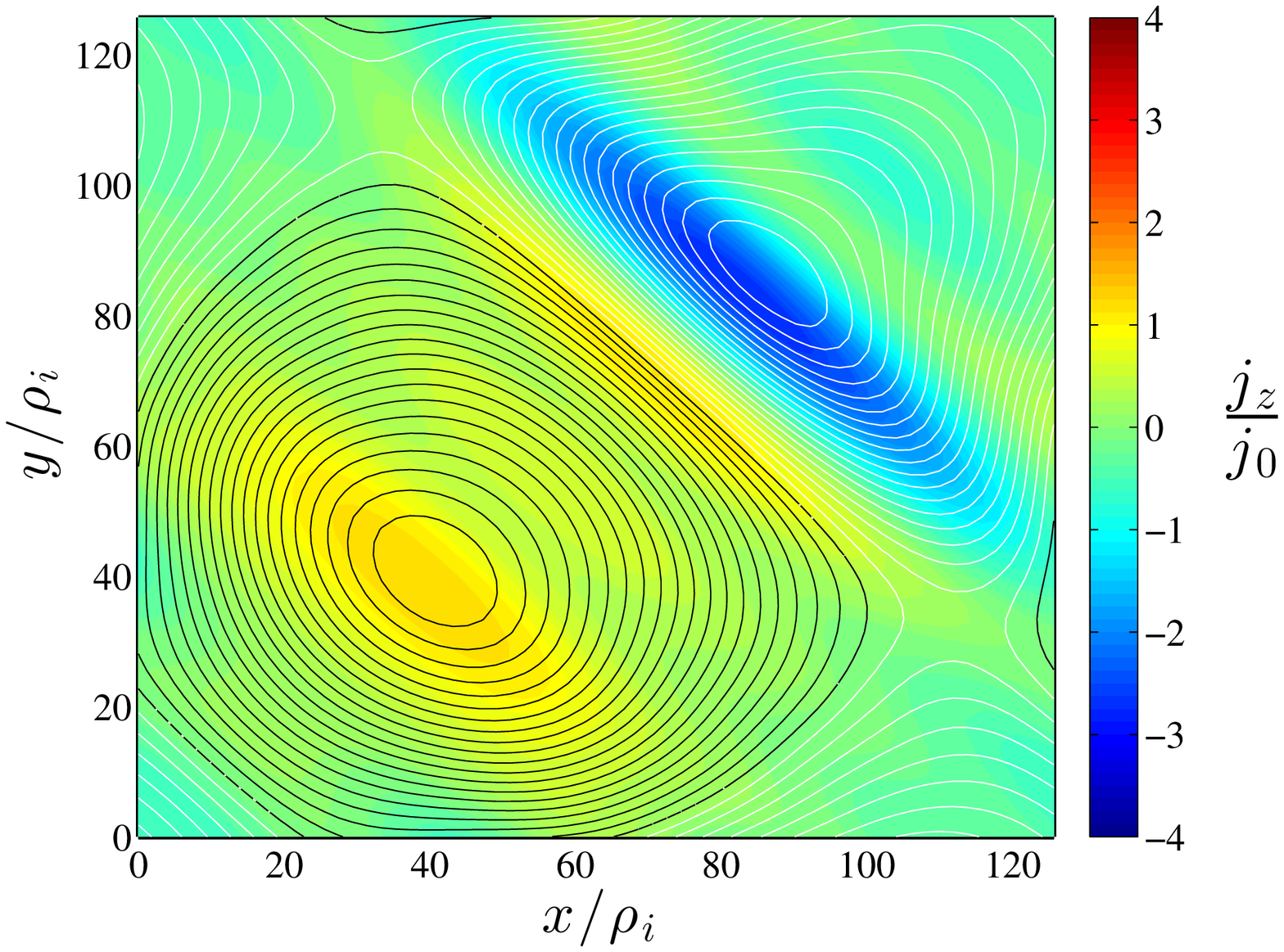}} \hfill}
\caption{ (a) Normalized current density $j_z/j_0$ (colorbar) and
  contours of parallel vector potential $A_\parallel$
  (positive--black, negative--white) in the perpendicular plane for an
  \Alfven wave collision simulation with $\chi=1$, showing the
  development of a current sheet (upper right quadrant, blue). (b)-(f)
  Filtering is used to remove all but the select number of Fourier
  modes shown in each panel. Constructive interference among just 12
  perpendicular Fourier modes (f) is sufficient to qualitatively
  reproduce the current sheet structure arising in the full simulation
  (a) of a strong \Alfven wave collision with $\chi=1$.
\label{fig:interfere}}
\end{figure*}

\section{Current Sheet Formation}
The primary numerical result reported here is the discovery that, as
the amplitudes of the colliding \Alfven waves increase to the strong
turbulence limit, $\chi \rightarrow 1$, the transient development of a
current sheet is a natural consequence. The current sheet resulting
from a nonlinear gyrokinetic simulation of a strong \Alfven wave
collision with $\chi=1$ is plotted in Fig.~\ref{fig:interfere}(a),
where the normalized parallel current $j_z/j_0$ (colorbar) and
positive (black) and negative (white) contours of the parallel vector
potential, $A_\parallel$, are plotted across a plane perpendicular to
the equilibrium magnetic field at $z=-L_\parallel/4$ and $t=1.30
\ T_A$, where the \Alfven wave period is $T_A=L_\parallel/V_A$ and
$j_0=n_0q_i v_{ti} L_\perp/L_\parallel$.

The current sheet in the upper right quadrant may be characterized by
its width $w$ and thickness $\delta$ in the perpendicular plane, its
length along the equilibrium magnetic field $l$, and its lifetime
$\tau$.  The width $w$, thickness $\delta$, and length $l$ are
determined using the full-width, half-maximum extent of the current
density in each of these directions. The lifetime $\tau$ is estimated
as the time over which its peak current density exceeds half the
global maximum in the simulation domain.  We find width $w \simeq
2L_\perp /3 \simeq 90 \rho_i$ and thickness $\delta \simeq 9 \rho_i$,
yielding an aspect ratio of $w/\delta \simeq 10$. The parallel length
is $l \simeq L_\parallel$, and the total lifetime from the beginning
of current sheet formation to the end of its decay is $\tau \simeq 3
T_A/4$.

So, what controls the self-consistently generated current sheet's
lifetime and morphology?  The length $l$ along the equilibrium
magnetic field and its width $w$ in the perpendicular plane are
determined by the parallel and perpendicular components of the
wavelengths of the original interacting \Alfven waves.  The lifetime
of this current sheet is related to the period of the original
(large-scale) \Alfven waves, a time much longer than the \Alfven
crossing time across the thickness $\delta$ of the current sheet.  The
thickness $\delta$ appears to approach the smallest resolvable
perpendicular scale in the simulation. However, in this particular
simulation, the thickness $\delta$ also happens to be approximately
the scale $k_\perp \rho_i \sim 1$ where the linear physics becomes
dispersive, in other words the thickness is coincident with the
perpendicular wavelength where the non-dispersive \Alfven waves
convert to dispersive kinetic \Alfven waves.  Further exploration is
required to determine whether the current sheet thickness is bounded
by the simulation resolution or by the decoupling of ions from the
electromagnetic fluctuations at $k_\perp \rho_i \sim 1$.

Note that plasma turbulence simulations ubiquitously show the
development of current sheets
\citep{Matthaeus:1980,Meneguzzi:1981,Wan:2012,Karimabadi:2013,TenBarge:2013a,Wu:2013,Zhdankin:2013}
which sometimes appear to persist for a long time relative to other
turbulent fluctuations on the scale of the thickness of the current
sheet. This mechanism for current sheet generation may explain why
these current sheets appear to persist for a long time because their
lifetime is governed by the interaction of much larger scale \Alfven
waves. The period of those large scale \Alfven waves is much longer
than the period of the turbulent fluctuations on the scale of the
current sheet thickness, resulting in a persistent current sheet.

\section{Physical mechanism of current sheet development}
The key conceptual result presented here is an explanation for this
current sheet development in terms of the nonlinear dynamics of strong
\Alfven wave collisions. In the weakly nonlinear limit, $\chi \ll 1$,
the nonlinearly generated \Alfven waves $\V{k}_3^\pm$ are the dominant
recipients of the energy transferred secularly to smaller scales,
where the self-consistently generated $\V{k}_2^{(0)}$ mode, which has
$k_z=0$, mediates the transfer. But, as the colliding \Alfven wave
amplitudes increase to the strongly nonlinear limit, $\chi \rightarrow
1$, the asymptotic expansion of the equations of evolution ceases to
be well-ordered, so higher-order terms---terms that can safely be
neglected in the weakly nonlinearly limit---begin to contribute
significantly. Nonetheless, the dominant pathway of the nonlinear
energy transfer mediated by self-consistently generated $k_z=0$ modes,
found in the weakly nonlinear limit, persists as one approaches and
reaches the strongly nonlinear limit, $\chi\rightarrow 1$, as shown in
Fig.~\ref{fig:chi_scan}.

This persistence of the nonlinear energy transfer mechanism is due to
the fact that the phases and amplitudes of all of the nonlinearly
generated modes are determined by the mathematical form of the
nonlinear term in eq.~(\ref{eq:elsasserpm}), even in the strongly
nonlinear limit. If the nonlinearly generated modes rise to sufficient
amplitudes, as occurs in the strongly nonlinear limit, they may
constructively interfere with the primary \Alfven waves to create a
coherent structure, in this case a current sheet.

A simple analogy for the development of a coherent structure by the
interference of a sum of sinusoidal modes is the square wave, given by
the infinite sum of Fourier modes, $f(x) = \sin x + \sin(3x)/3 +
\sin(5x)/5 + \ldots$.  Current sheet development in plasma turbulence
is similar, but involves the sum of Fourier modes spanning the
two-dimensional plane perpendicular to the local mean magnetic field,
as depicted in Fig.~\ref{fig:modes_o5}. The fluctuations arising from
the turbulent cascade, although they may appear random, in fact have
phase and amplitude relationships determined by the nonlinear terms in
the governing nonlinear equations of evolution. It is these
nonlinearly determined phase and amplitude relationships that give
rise to the coherent structures that arise from what may appear to be
random turbulent fluctuations.  A consequence of this insight is that
the current sheet development can be predicted analytically, as is
demonstrated below by a direct comparison between the analytical
calculation and a nonlinear numerical simulation for the moderately
nonlinear case of $\chi=1/2$.

\begin{figure}
\begin{center}
  \resizebox{3.2in}{!}{\includegraphics{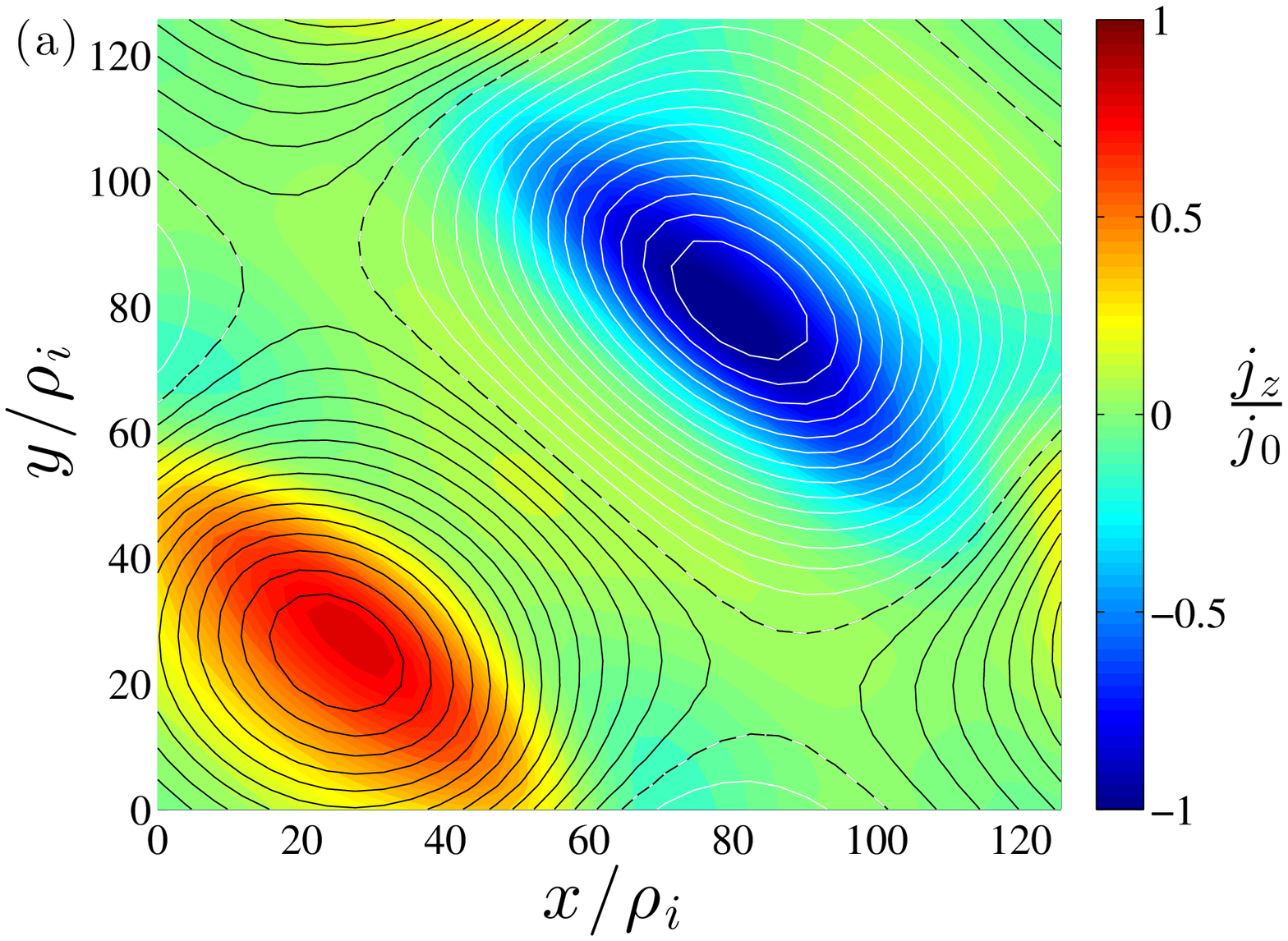}}
  \resizebox{3.2in}{!}{\includegraphics{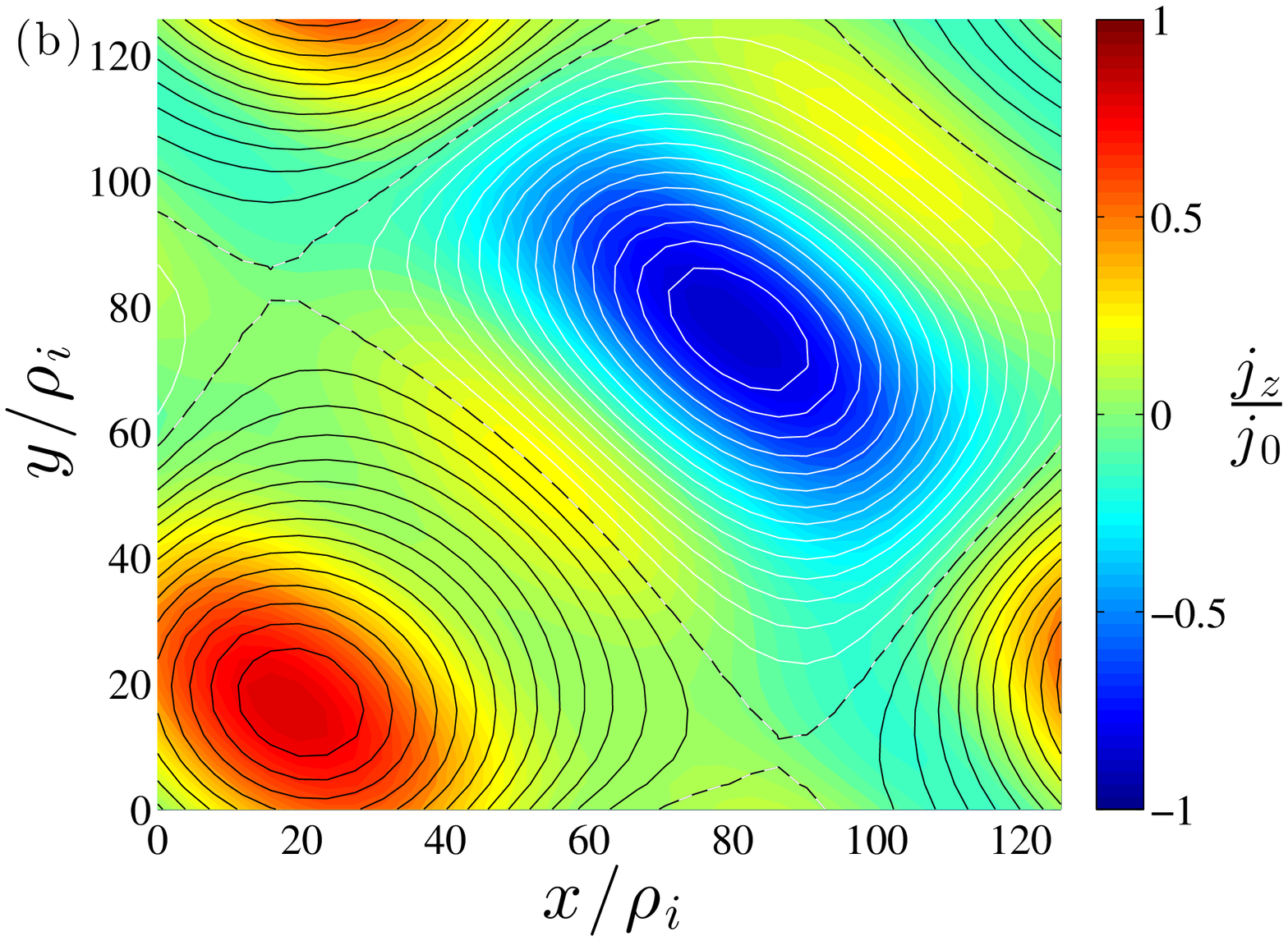}}
  \end{center}
\caption{(a) Normalized current $j_z/j_0$ (colorbar) and contours of
  parallel vector potential $A_\parallel$ (positive--black,
  negative--white) in the perpendicular plane for an \Alfven wave
  collision simulation with $\chi=1/2$, showing the tendency for the
  current to elongate into a sheet-like coherent structure. (b)
  Analytical prediction of the current sheet development for the same
  \Alfven wave collision.
\label{fig:predict}}
\end{figure}

In Fig.~\ref{fig:predict}(a), we plot the normalized parallel current
$j_z/j_0$ across the perpendicular plane at $t=3.64 \ T_A$ for the
$\chi=1/2$ simulation. Although the current sheet formation here is
not as rapid or intense as the $\chi=1$ case, the tendency for the
current to elongate into a sheet-like coherent structure is
apparent. Using the analytical solution for the nonlinear evolution of
\Alfven wave collisions presented in \citet{Howes:2013a}, we compute a
first-principles, analytical prediction of the current sheet
development for this $\chi=1/2$ simulation, shown in
Fig.~\ref{fig:predict}(b). For this analytical prediction, we have
included only the five lowest order modes shown in Fig.~1:
$\V{k}_1^+$, $\V{k}_1^-$, $\V{k}_2^{(0)}$, $\V{k}_3^+ $, and
$\V{k}_3^-$. This demonstration that the current sheet development is
a result of interference between just five complex Fourier modes in
the perpendicular plane, and can be predicted analytically from a
rigorous first-principles calculation, supports the hypothesis that
interference between the primary and nonlinearly generated Fourier
modes is responsible for the development of current sheets in \Alfven
wave collisions.

But the qualitative properties of the nonlinear energy transfer imply
a further major simplification of the nonlinear evolution.  For the
particular wavevectors specified for the two primary
counterpropagating \Alfven waves, $\V{k}_1^\pm$, the mathematical form
of the nonlinear term in eq.~(\ref{eq:elsasserpm}) fixes the value of
$k_z$ for all nonlinearly generated modes to be constant along lines
of slope 1 (thin diagonal arrows) on the $(k_x,k_y)$ plane in
Fig.~\ref{fig:modes_o5}.  Since the energy transfer is mediated by
self-consistently generated $k_z=0$ modes, energy is preferentially
transferred to Fourier modes falling along the $k_z=+k_\parallel$,
$k_z=0$, and $k_z=-k_\parallel$ lines \citep{Howes:2013a}, so it is
just these relatively few modes that govern the current sheet
development. In Fig.~\ref{fig:interfere}(b)--(f), we plot a
successively increasing number of modes just along these three
diagonal lines, filtering out all other modes from the $\chi=1$
simulation: (b) just the two primary \Alfven waves
$(k_x/k_\perp,k_y/k_\perp)=(0,1)$ and $(1,0)$; (c) adding the
self-consistently generated $k_z=0$ mode at $(1,1)$ for 3 total modes;
(d) adding $(2,1)$, $(2,2)$, and $(1,2)$ for 6 modes; (e) adding
$(3,2)$, $(3,3)$, and $(2,3)$ for 9 modes; and (f) adding $(4,3)$,
$(4,4)$, and $(3,4)$ for 12 total modes.  The successive addition of
these modes in this figure confirms that constructive interference
among these modes leads to the observed current sheet structure. By
comparing the results of the full simulation in (a) to the plot with
just 12 modes in (f)---only 1.3\% of the 945 possible complex Fourier
modes in the simulation---the structure of the current sheet from the
full simulation is quite accurately reproduced, showing only minor
quantitative differences.

\begin{figure}
\begin{center}
\resizebox{3.5in}{!}{\includegraphics{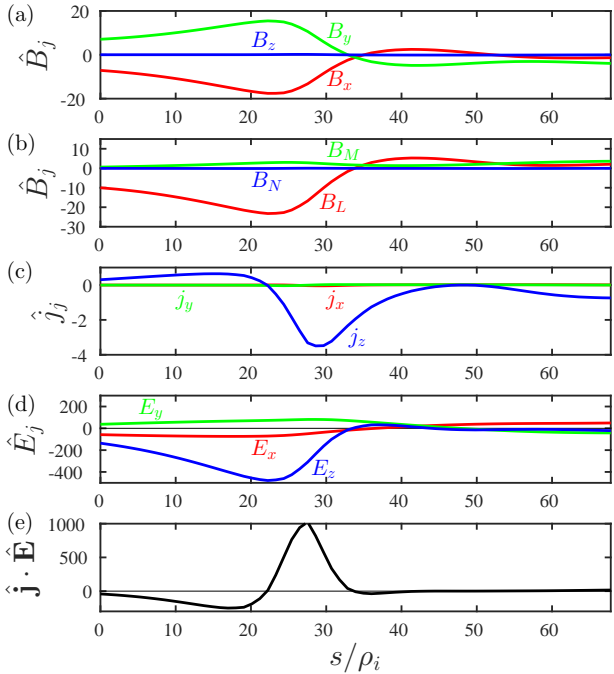}}
\end{center}
\caption{Profiles of normalized magnetic field in (a) simulation
  coordinates and (b) minimum variance coordinates, (c) current, (d)
  electric field, and (e) $\V{j} \cdot \V{E}$ vs.~normalized distance
  $s/\rho_i$ along the trajectory shown in Fig.~\ref{fig:interfere}(a)
  (black line).
\label{fig:slices}}
\end{figure}

\section{Comparison to spacecraft observations}
Do the current sheets generated by strong \Alfven wave collisions have
similar properties to those measured in space plasmas? To address this
important question, we sample the current sheet in our simulation
along the trajectory given by the black line in
Fig.~\ref{fig:interfere}(a). In Fig.~\ref{fig:slices}, we plot the
resulting profiles of the normalized magnetic field $\hat{B}_j=(\delta
B_j/B_0)(L_\parallel/L_\perp)$ in both simulation $(x,y,z)$ and
minimum variance $(L,M,N)$ coordinates, normalized current
$\hat{j}_j=j_j/j_0$, normalized electric field $\hat{E}_j=(c E_j)/(v_A
B_0)(L_\parallel/L_\perp)$, and normalized $\hat{\V{j}} \cdot
\hat{\V{E}}$. These profiles compare favorably with published
measurements of the magnetic field variation in minimum variance
coordinates and of $\V{j} \cdot\V{E}$ across current sheets from
turbulence in the magnetosheath \citep{Retino:2007,Sundkvist:2007} and
solar wind \citep{Perri:2012a}. This preliminary comparison motivates
future efforts to make a more detailed statistical comparison of the
properties of current sheets arising in strong \Alfven wave collisions
with those measured by spacecraft missions.

\section{Conclusion} 
The nonlinear dynamics of strong \Alfven wave collisions provides a
natural explanation for the ubiquitous development of current sheets
in plasma turbulence. The discovery that current sheets arise
transiently through constructive interference among the primary waves
and nonlinearly generated modes provides valuable insight into the
physical mechanisms by which the turbulent fluctuations are
damped. Because the dominant constructively interfering modes are
$k_z=+k_\parallel$ ($k_z=-k_\parallel$) \Alfven waves that propagate
in the $+\zhat$ ($-\zhat$) direction and $k_z=0$ modes nonlinearly
generated by the interactions between these counterpropagating waves,
collisionless damping of the constituent \Alfven waves via the Landau
resonance with protons and electrons is expected to play an important
role in the dissipation of the current sheets, as previously suggested
\citep{TenBarge:2013a}.  Of course, other mechanisms exist that can
produce current sheets, such as particular flow and magnetic field
geometries, like the Orszag-Tang vortex. A final
question that remains to be answered is whether such alternative
mechanisms play any role in the development of current sheets in solar
wind turbulence, or are strong \Alfven wave collisions sufficient to
account for all current sheets observed in the turbulent solar wind?

This work was supported by NSF grant PHY-10033446, NSF CAREER Award
AGS-1054061, and NASA grant NNX10AC91G. This work used the Extreme
Science and Engineering Discovery Environment (XSEDE), which is
supported by National Science Foundation grant number ACI-1053575.



\begin{thebibliography}{50}
\expandafter\ifx\csname natexlab\endcsname\relax\def\natexlab#1{#1}\fi

\bibitem[{{Abel} {et~al.}(2008){Abel}, {Barnes}, {Cowley}, {Dorland}, \&
  {Schekochihin}}]{Abel:2008}
{Abel}, I.~G., {Barnes}, M., {Cowley}, S.~C., {Dorland}, W., \& {Schekochihin},
  A.~A. 2008, Phys.~Plasmas, 15, 122509

\bibitem[{{Antiochos}(1987)}]{Antiochos:1987}
{Antiochos}, S.~K. 1987, Astrophys.~J., 312, 886

\bibitem[{{Barnes} {et~al.}(2009){Barnes}, {Abel}, {Dorland}, {Ernst},
  {Hammett}, {Ricci}, {Rogers}, {Schekochihin}, \& {Tatsuno}}]{Barnes:2009}
{Barnes}, M., {et~al.} 2009, Phys.~Plasmas, 16, 072107

\bibitem[{{Biskamp} \& {M{\"u}ller}(2000)}]{Biskamp:2000}
{Biskamp}, D., \& {M{\"u}ller}, W.-C. 2000, Phys.~Plasmas, 7, 4889

\bibitem[{{Biskamp} \& {Welter}(1989)}]{Biskamp:1989}
{Biskamp}, D., \& {Welter}, H. 1989, Phys.~Fluids B, 1, 1964

\bibitem[{{Boldyrev}(2006)}]{Boldyrev:2006}
{Boldyrev}, S. 2006, Phys.~Rev.~Lett., 96, 115002

\bibitem[{{Boldyrev} {et~al.}(2011){Boldyrev}, {Perez}, {Borovsky}, \&
  {Podesta}}]{Boldyrev:2011}
{Boldyrev}, S., {Perez}, J.~C., {Borovsky}, J.~E., \& {Podesta}, J.~J. 2011,
  Astrophys.~J.~Lett., 741, L19

\bibitem[{{Borovsky}(2008)}]{Borovsky:2008}
{Borovsky}, J.~E. 2008, J.~Geophys.~Res., 113, 8110

\bibitem[{{Borovsky}(2010)}]{Borovsky:2010}
---. 2010, Phys.~Rev.~Lett., 105, 111102

\bibitem[{{Borovsky} \& {Denton}(2011)}]{Borovsky:2011}
{Borovsky}, J.~E., \& {Denton}, M.~H. 2011, Astrophys.~J.~Lett., 739, L61

\bibitem[{{Cowley} {et~al.}(1997){Cowley}, {Longcope}, \&
  {Sudan}}]{Cowley:1997}
{Cowley}, S.~C., {Longcope}, D.~W., \& {Sudan}, R.~N. 1997, Phys.~Rep., 283,
  227

\bibitem[{{Frieman} \& {Chen}(1982)}]{Frieman:1982}
{Frieman}, E.~A., \& {Chen}, L. 1982, Phys.~Fluids, 25, 502

\bibitem[{{Galtier} {et~al.}(2000){Galtier}, {Nazarenko}, {Newell}, \&
  {Pouquet}}]{Galtier:2000}
{Galtier}, S., {Nazarenko}, S.~V., {Newell}, A.~C., \& {Pouquet}, A. 2000,
  J.~Plasma Phys., 63, 447

\bibitem[{Goldreich \& Sridhar(1995)}]{Goldreich:1995}
Goldreich, P., \& Sridhar, S. 1995, Astrophys.~J., 438, 763

\bibitem[{{Greco} {et~al.}(2008){Greco}, {Chuychai}, {Matthaeus}, {Servidio},
  \& {Dmitruk}}]{Greco:2008}
{Greco}, A., {Chuychai}, P., {Matthaeus}, W.~H., {Servidio}, S., \& {Dmitruk},
  P. 2008, Geophys.~Res.~Lett., 35, 19111

\bibitem[{Howes(2015)}]{Howes:2015b}
Howes, G.~G. 2015, Philosophical Transactions of the Royal Society of London A:
  Mathematical, Physical and Engineering Sciences, 373, 20140145

\bibitem[{{Howes} {et~al.}(2006){Howes}, {Cowley}, {Dorland}, {Hammett},
  {Quataert}, \& {Schekochihin}}]{Howes:2006}
{Howes}, G.~G., {Cowley}, S.~C., {Dorland}, W., {Hammett}, G.~W., {Quataert},
  E., \& {Schekochihin}, A.~A. 2006, Astrophys.~J., 651, 590

\bibitem[{{Howes} {et~al.}(2012){Howes}, {Drake}, {Nielson}, {Carter},
  {Kletzing}, \& {Skiff}}]{Howes:2012b}
{Howes}, G.~G., {Drake}, D.~J., {Nielson}, K.~D., {Carter}, T.~A., {Kletzing},
  C.~A., \& {Skiff}, F. 2012, Phys.~Rev.~Lett., 109, 255001

\bibitem[{{Howes} \& {Nielson}(2013)}]{Howes:2013a}
{Howes}, G.~G., \& {Nielson}, K.~D. 2013, Phys.~Plasmas, 20, 072302

\bibitem[{Iroshnikov(1963)}]{Iroshnikov:1963}
Iroshnikov, R.~S. 1963, Astron. Zh., 40, 742, {English} Translation: Sov.
  Astron., 7 566 (1964)

\bibitem[{{Karimabadi} {et~al.}(2013){Karimabadi}, {Roytershteyn}, {Wan},
  {Matthaeus}, {Daughton}, {Wu}, {Shay}, {Loring}, {Borovsky}, {Leonardis},
  {Chapman}, \& {Nakamura}}]{Karimabadi:2013}
{Karimabadi}, H., {et~al.} 2013, Phys.~Plasmas, 20, 012303

\bibitem[{Kraichnan(1965)}]{Kraichnan:1965}
Kraichnan, R.~H. 1965, Phys.~Fluids, 8, 1385

\bibitem[{{Longcope} \& {Strauss}(1994)}]{Longcope:1994}
{Longcope}, D.~W., \& {Strauss}, H.~R. 1994, Astrophys.~J., 437, 851

\bibitem[{{Matthaeus} \& {Montgomery}(1980)}]{Matthaeus:1980}
{Matthaeus}, W.~H., \& {Montgomery}, D. 1980, Annals of the New York Academy of
  Sciences, 357, 203

\bibitem[{{Meneguzzi} {et~al.}(1981){Meneguzzi}, {Frisch}, \&
  {Pouquet}}]{Meneguzzi:1981}
{Meneguzzi}, M., {Frisch}, U., \& {Pouquet}, A. 1981, Phys.~Rev.~Lett., 47,
  1060

\bibitem[{{Merrifield} {et~al.}(2005){Merrifield}, {M{\"u}ller}, {Chapman}, \&
  {Dendy}}]{Merrifield:2005}
{Merrifield}, J.~A., {M{\"u}ller}, W.-C., {Chapman}, S.~C., \& {Dendy}, R.~O.
  2005, Phys.~Plasmas, 12, 022301

\bibitem[{{Ng} \& {Bhattacharjee}(1996)}]{Ng:1996}
{Ng}, C.~S., \& {Bhattacharjee}, A. 1996, Astrophys.~J., 465, 845

\bibitem[{{Nielson} {et~al.}(2013){Nielson}, {Howes}, \&
  {Dorland}}]{Nielson:2013a}
{Nielson}, K.~D., {Howes}, G.~G., \& {Dorland}, W. 2013, Physics of Plasmas,
  20, 072303

\bibitem[{{Numata} {et~al.}(2010){Numata}, {Howes}, {Tatsuno}, {Barnes}, \&
  {Dorland}}]{Numata:2010}
{Numata}, R., {Howes}, G.~G., {Tatsuno}, T., {Barnes}, M., \& {Dorland}, W.
  2010, J.~Comp.~Phys., 229, 9347

\bibitem[{{Osman} {et~al.}(2014){Osman}, {Matthaeus}, {Gosling}, {Greco},
  {Servidio}, {Hnat}, {Chapman}, \& {Phan}}]{Osman:2014b}
{Osman}, K.~T., {Matthaeus}, W.~H., {Gosling}, J.~T., {Greco}, A., {Servidio},
  S., {Hnat}, B., {Chapman}, S.~C., \& {Phan}, T.~D. 2014, Phys.~Rev.~Lett.,
  112, 215002

\bibitem[{{Osman} {et~al.}(2011){Osman}, {Matthaeus}, {Greco}, \&
  {Servidio}}]{Osman:2011}
{Osman}, K.~T., {Matthaeus}, W.~H., {Greco}, A., \& {Servidio}, S. 2011,
  Astrophys.~J.~Lett., 727, L11+

\bibitem[{{Osman} {et~al.}(2012){Osman}, {Matthaeus}, {Wan}, \&
  {Rappazzo}}]{Osman:2012a}
{Osman}, K.~T., {Matthaeus}, W.~H., {Wan}, M., \& {Rappazzo}, A.~F. 2012,
  Phys.~Rev.~Lett., 108, 261102

\bibitem[{{Parker}(1972)}]{Parker:1972}
{Parker}, E.~N. 1972, Astrophys.~J., 174, 499

\bibitem[{{Perri} {et~al.}(2012){Perri}, {Goldstein}, {Dorelli}, \&
  {Sahraoui}}]{Perri:2012a}
{Perri}, S., {Goldstein}, M.~L., {Dorelli}, J.~C., \& {Sahraoui}, F. 2012,
  Phys.~Rev.~Lett., 109, 191101

\bibitem[{{Pouquet}(1978)}]{Pouquet:1978}
{Pouquet}, A. 1978, J.~Fluid Mech., 88, 1

\bibitem[{{Priest}(1985)}]{Priest:1985}
{Priest}, E.~R. 1985, Reports on Progress in Physics, 48, 955

\bibitem[{{Retin{\`o}} {et~al.}(2007){Retin{\`o}}, {Sundkvist}, {Vaivads},
  {Mozer}, {Andr{\'e}}, \& {Owen}}]{Retino:2007}
{Retin{\`o}}, A., {Sundkvist}, D., {Vaivads}, A., {Mozer}, F., {Andr{\'e}}, M.,
  \& {Owen}, C.~J. 2007, Nature Physics, 3, 236

\bibitem[{{She} {et~al.}(1990){She}, {Jackson}, \& {Orszag}}]{She:1990}
{She}, Z.-S., {Jackson}, E., \& {Orszag}, S.~A. 1990, Nature, 344, 226

\bibitem[{{Spangler}(1999)}]{Spangler:1999}
{Spangler}, S.~R. 1999, Astrophys.~J., 522, 879

\bibitem[{{Sridhar} \& {Goldreich}(1994)}]{Sridhar:1994}
{Sridhar}, S., \& {Goldreich}, P. 1994, Astrophys.~J., 432, 612

\bibitem[{{Sundkvist} {et~al.}(2007){Sundkvist}, {Retin{\`o}}, {Vaivads}, \&
  {Bale}}]{Sundkvist:2007}
{Sundkvist}, D., {Retin{\`o}}, A., {Vaivads}, A., \& {Bale}, S.~D. 2007,
  Phys.~Rev.~Lett., 99, 025004

\bibitem[{{TenBarge} \& {Howes}(2013)}]{TenBarge:2013a}
{TenBarge}, J.~M., \& {Howes}, G.~G. 2013, Astrophys.~J.~Lett., 771, L27

\bibitem[{{Uritsky} {et~al.}(2010){Uritsky}, {Pouquet}, {Rosenberg}, {Mininni},
  \& {Donovan}}]{Uritsky:2010}
{Uritsky}, V.~M., {Pouquet}, A., {Rosenberg}, D., {Mininni}, P.~D., \&
  {Donovan}, E.~F. 2010, Phys.~Rev.~E, 82, 056326

\bibitem[{{van Ballegooijen}(1985)}]{vanBallegooijen:1985}
{van Ballegooijen}, A.~A. 1985, Astrophys.~J., 298, 421

\bibitem[{{Wan} {et~al.}(2012){Wan}, {Matthaeus}, {Karimabadi}, {Roytershteyn},
  {Shay}, {Wu}, {Daughton}, {Loring}, \& {Chapman}}]{Wan:2012}
{Wan}, M., {et~al.} 2012, Phys.~Rev.~Lett., 109, 195001

\bibitem[{{Wang} {et~al.}(2013){Wang}, {Tu}, {He}, {Marsch}, \&
  {Wang}}]{Wang:2013}
{Wang}, X., {Tu}, C., {He}, J., {Marsch}, E., \& {Wang}, L. 2013,
  Astrophys.~J.~Lett., 772, L14

\bibitem[{{Wu} {et~al.}(2013){Wu}, {Perri}, {Osman}, {Wan}, {Matthaeus},
  {Shay}, {Goldstein}, {Karimabadi}, \& {Chapman}}]{Wu:2013}
{Wu}, P., {et~al.} 2013, Astrophys.~J.~Lett., 763, L30

\bibitem[{{Zhdankin} {et~al.}(2012){Zhdankin}, {Boldyrev}, {Mason}, \&
  {Perez}}]{Zhdankin:2012}
{Zhdankin}, V., {Boldyrev}, S., {Mason}, J., \& {Perez}, J.~C. 2012,
  Phys.~Rev.~Lett., 108, 175004

\bibitem[{{Zhdankin} {et~al.}(2013){Zhdankin}, {Uzdensky}, {Perez}, \&
  {Boldyrev}}]{Zhdankin:2013}
{Zhdankin}, V., {Uzdensky}, D.~A., {Perez}, J.~C., \& {Boldyrev}, S. 2013,
  Astrophys.~J., 771, 124

\bibitem[{{Zweibel} \& {Li}(1987)}]{Zweibel:1987}
{Zweibel}, E.~G., \& {Li}, H.-S. 1987, Astrophys.~J., 312, 423

\end{thebibliography}

\end{document}